# Lattice Thermal Conductivity of Organic-Inorganic Hybrid Perovskite $CH_3NH_3PbI_3$

Xin Qian, Xiaokun Gu and Ronggui Yang[*]

Department of Mechanical Engineering & Materials Science and Engineering Program, University of Colorado, Boulder, Colorado 80309-0427, USA

[*]E-mail: Ronggui.Yang@Colorado.edu

**Abstract**

Great success has been achieved in improving the photovoltaic energy conversion efficiency of the organic-inorganic hybrid perovskite-based solar cells, but with very limited knowledge on the thermal transport in hybrid perovskites, which could affect the device lifetime and stability. Based on the potential field derived from the density functional theory calculations, we studied the lattice thermal conductivity of the hybrid halide perovskite $CH_3NH_3PbI_3$ using equilibrium molecular dynamics simulations. Temperature-dependent thermal conductivity is reported from 160 K to 400 K, which covers the tetragonal phase (160 – 330 K) and the pseudocubic phase (>330 K). A very low thermal conductivity (0.59 W/m·K) is found in the tetragonal phase at room temperature, whereas a much higher thermal conductivity is found in the pseudocubic phase (1.80 W/m·K at 330K). The low group velocity of acoustic phonons and the strong anharmonicity are found responsible for the relatively low thermal conductivity of the tetragonal $CH_3NH_3PbI_3$.

The hybrid perovskite $CH_3NH_3PbI_3$ (hereafter called as $MAPbI_3$ where MA denotes the $CH_3NH_3$ cation) has showed great potential in photovoltaic applications due to its high absorption coefficient, long diffusion length of charge carriers and easiness of processing.[1-6] Although great success has been achieved in improving the photovoltaic energy conversion efficiency,[6-10] heat dissipation and thermal stability of $MAPbI_3$ remain to be a concern, which affects the device lifetime and efficiency.[11-13] Therefore, understanding the thermal transport in the $MAPbI_3$ is important for well-designed hybrid perovskite-based energy conversion devices. The knowledge on thermal properties of $MAPbI_3$ has been limited with only one group reported experimental measurement of thermal conductivity below room temperature.[14] In solar cell applications, the device temperature could reach above 350 K under continuous solar illumination,[12] which would invoke a phase transition of $MAPbI_3$ from the tetragonal phase at room temperature to its pseudocubic polymorph at higher temperature.[15]

Besides the thermal stability concerns in solar cells, there are also great interests in potential thermoelectric applications of hybrid perovskites due to the reported low thermal conductivity[14] and the high mobility of charge carriers[16]. Recent studies reported that thermoelectric figure of merit ZT can reach close to unity by increasing charge carrier concentration through photo-induced or chemical doping.[17, 18]

In this letter, we studied the lattice thermal conductivity of the hybrid perovskite $MAPbI_3$ from 160 K to 400 K. As temperature increases, $MAPbI_3$ exhibits a phase transition from the tetragonal (160 – 330 K) to the pseudocubic phase (>330 K).[15] For each phase, an empirical potential field is developed based on density functional theory (DFT) calculations [19], and molecular dynamics (MD) simulations are then performed to extract the lattice thermal conductivity.[20]

The DFT calculations are first performed to develop the empirical potential field. We use Quantum Espresso package for all DFT calculations[21], based on Perdew-Burke-Ernzerhof functionals optimized for solids (PBEsol)[22] and the projector augmented-wave (PAW) method[23]. Valence electrons include not only the s- and p- orbitals in the outer electron shell for all atoms, but also the 5d orbitals in Pb atoms. At first, self-consistent field calculations are performed to test the convergence of ground state energy with respect to Monkhost−Pack mesh (k-mesh) density and the energy cutoff. The convergence of the ground-state energy was reached when using a 60 Ry energy cutoff, and a k-mesh of 6×6×4 for the tetragonal phase and a k-mesh of 8×8×8 for the pseudocubic phase, respectively (see Figure S1 and Figure S2 of Supplementary Information). The crystal structure is then relaxed using Broyden–Fletcher–Goldfarb–Shanno algorithm[24] until the interatomic forces is smaller than 5meV/Å.[17] Relaxed crystal structures of the tetragonal and the pseudocubic phases of MAPbI$_3$ are shown in Figure 1. Table 1 shows that the maximum error of lattice constants between our DFT calculation and the X-ray diffraction (XRD) experiments[15] is within 2%, which is acceptable for DFT calculations.[25] Bader charge analysis[26] is then performed to obtain the atomic charges for the Coulombic interactions (see Table 2) of the potential field. We have used a dense (60×60×60) fast Fourier transform mesh (FFT mesh) for wave functions to reconstruct the density distribution of all electrons. We observed that further doubling the FFT grid density would only result in 2% change of the atomic charges.

Based on the relaxed crystal structure, *ab initio* molecular dynamics (AIMD) simulations are performed to obtain the relationship between interatomic forces and atomic displacements, which is then fitted to develop the potential field. We used a 2×2×2 supercell at 300 K for the tetragonal phase and 3×3×3 supercell at 400 K for the pseudocubic phase. The duration of AIMD simulation is 100 fs with a 1 fs time step. This AIMD simulation essentially randomly samples the slope of the potential energy surface at different atomic positions as a function of time $t$: $\boldsymbol{F}_{AIMD}(i, t) = -\nabla E_{AIMD}(r_i(t))$, where $\boldsymbol{F}_{AIMD}$ and $E_{AIMD}$ is the force and potential energy predicted by AIMD, and $r_i(t)$ is the position trajectory of atom $i$. The objective of developing a potential field is to match the energy surface curvature by DFT calculations with the assigned functional forms which contains the unknown parameters $\boldsymbol{P}$. For any set of parameters $\boldsymbol{P}$, the force imposed on atom $i$ is also calculated as the gradient of the potential energy surface: $\boldsymbol{F}_V(\boldsymbol{P}; i, t) = -\nabla E_V(\boldsymbol{P}; r_i(t))$, where the potential surface $E_V$ is the summation of all interaction terms listed in Table 2. The error of the developed potential field is defined as the squared deviation between $\boldsymbol{F}_V$ and $\boldsymbol{F}_{AIMD}$ summed over all time steps for all atoms:

$$W(\boldsymbol{P}) = \frac{1}{N_t N} \sum_{t=1}^{N_t} \sum_i |\boldsymbol{F}_V(\boldsymbol{P}; i, t) - \boldsymbol{F}_{AIMD}(i, t)|^2 \tag{1}$$

where $N$ is the number of atoms in the simulation cell and $N_t$ is number of time step in the AIMD simulation.[27] The parameters $\boldsymbol{P}$ is adjusted using the quasi-Newton algorithm[28] to seek the minimum value of the force deviation function $W$. The parameters listed in Table 2 are found to give the minimal $W$ for the tetragonal and pseudocubic MAPbI$_3$. To evaluate the accuracy of the empirical potential fields, we have calculated the phonon density of states of MAPbI$_3$ for both phases using MD simulations (see Figure S4 in Supplementary Information), which shows a reasonable agreement with the vibration spectrum from DFT calculations by Brivio et al.[29]

With the empirical potential field developed, equilibrium molecular dynamics (EMD) simulations are then performed using LAMMPS package[30] to extract the thermal conductivity of MAPbI$_3$ from 160 K to 400 K, where a phase transition occurs at 330K. We use a 1fs time step to integrate Newton's equation of atomic motion by Verlet algorithm. At each specific temperature, the thermal conductivity $k$ of MAPbI$_3$ is extracted using the Green-Kubo relation:

$$k = \frac{1}{3Vk_BT^2} \int_0^\infty \langle \dot{\boldsymbol{w}}(0) \cdot \dot{\boldsymbol{w}}(t) \rangle dt \qquad (2)$$

where $k_B$ is the Boltzmann constant, $T$ is the temperature, $V$ is the volume of the simulation cell, $t$ is the correlation time and $\langle \dot{\boldsymbol{w}}(0) \cdot \dot{\boldsymbol{w}}(t) \rangle$ denotes the averaged heat current autocorrelation function (HCACF). The heat current $\dot{\boldsymbol{w}}$ is defined as the time derivative of the first-order energy moment in respect to the atomic position $\boldsymbol{r}_i$:

$$\dot{\boldsymbol{w}} = \frac{d}{dt} \sum_i E_i \boldsymbol{r}_i \qquad (3)$$

where $E_i$ is the total energy of atom $i$.

At the beginning of MD simulation, the atom position and lattice constants of MAPbI$_3$ are adjusted using conjugate gradient algorithm until the potential energy of the system is minimized. This structure optimization step in MD simulations is performed to make sure that the atoms are at the equilibrium positions in MD simulations, because the potential field developed does not perfectly match the energy surface from DFT calculation due to the limitation of the functional forms assigned.[19] As expected, the lattice constants are slightly changed after the MD structure optimization, but still within 2.5% error compared with the XRD results (see Table 1). After the structure optimization, the system is thermalized using NVT ensemble (canonical ensemble) for 200ps where the temperature fluctuation is minimized to be around 5 K. The simulation system is then switched to NVE ensemble (microcanonical ensemble) to run for 20 ns. During this time range, the HCACF for each phase is sampled every 2fs and plotted as shown in Figure 2. The HCACF vanishes, and the thermal conductivity calculated by equation (2) is converged when the correlation time reaches 50 ps for the tetragonal phase and 200 ps for the pseudocubic phase. The thermal conductivity is therefore calculated every 50 ps for the tetragonal phase and 200 ps for the pseudocubic phase using equation (2), and then averaged during the entire simulation time. To further obtain a better ensemble average, five independent runs are performed with different initial atomic velocities at each temperature we studied, and the final thermal conductivity is averaged over all obtained results.

Figure 3 shows the temperature-dependent thermal conductivity of MAPbI$_3$ from 160 K to 400 K, compared with the measurement by Pisoni et al.[14] The very low measured thermal conductivity of the tetragonal MAPbI$_3$ from 160 K to 330 K is confirmed by our MD simulation. At room temperature, the calculated thermal conductivity is 0.59 W/m·K, within 20% error compared with the experimental result of 0.50 W/m·K.[14] The thermal conductivity of MAPbI$_3$ shows a sudden jump from 0.57 W/m·K to 1.80 W/m·K at 330 K when phase-transition occurs from the tetragonal structure to the cubic structure. From the simple kinetic theory, the lattice thermal conductivity is determined by $k \sim \frac{1}{3} C v_g^2 \tau$ where $C$ is the volumetric heat capacity, $v_g$ is the phonon group velocity and $\tau$ is the phonon lifetime. We have thus evaluated and compared these three parameters for both the tetragonal and pseudocubic phases to explain the sudden increase in thermal conductivity of MAPbI$_3$ at the phase-transition temperature from the tetragonal to the pseudocubic structure.

First of all, the sudden increase in thermal conductivity at phase transition temperature should not be attributed to the change in the volumetric heat capacity. Based on the molar heat capacity ($C_m$) measurement by Onoda-Yamamuro et al.[31] and the density ($\rho$) of MAPbI$_3$ [15, 32], we calculated the volumetric heat capacity $C = C_m\rho/M$, where the $M$ is the molar mass of the MAPbI$_3$ (0.62 kg/mol). As shown Figure 3, the volumetric heat capacity of the pseudocubic MAPbI$_3$ is even slightly smaller than its tetragonal counterpart except for the peak at phase change temperature 330 K. Therefore, the heat capacity should not be responsible for the predicted sudden increase in thermal conductivity from the tetragonal phase to the pseudocubic phase.

We then evaluate the role of phonon lifetime and group velocity in the thermal conductivity difference between the two phases by calculating the spectral energy density (SED) [33, 34] distribution in the momentum-frequency domain. As shown in the Figure 4a and b, the location of SED peaks (bright colors) indicates the phonon dispersion and the bandwidth of the branches are related to the phonon lifetime. The smaller the broadening of the SED peak, the longer the phonon lifetime.[31, 35] In the tetragonal phase of MAPbI$_3$, the optical phonon branches (see Figure 4a) are significantly broadened, indicating very short optical phonon lifetimes. The line shape of the optical braches are still distinguishable in the pseudocubic phase, meaning that the optical phonons in the pseudocubic structure have much longer lifetimes (see Figure 4b). As an example, we plotted the SED peaks of the two phases at wavevector (0.1,0,0) in Figure 4c. The SED distribution of the tetragonal phase is very diffuse. In contrast, the SED peaks of optical braches in the pseudocubic phase can still be easily identified. We can conclude that the lower thermal conductivity in the tetragonal phase of MAPbI$_3$ is partly due to the shorter lifetimes of the optical phonon branches.

To calculate acoustic phonon velocity, we use polynomial interpolation to connect the points with the local maximum SEDs, and then take the first-order derivative at Brillouin zone center. As shown in Table 3, the group velocities of acoustic phonons in tetragonal MAPbI$_3$ are much smaller than that in the pseudocubic phase, which also results in the lower thermal conductivity of the tetragonal MAPbI$_3$. This is consistent with the observation that the tetragonal MAPbI$_3$ has much lower Young's modulus and shear modulus than the tetragonal phase.[36]

In summary, we have calculated the lattice thermal conductivity of the hybrid perovskite MAPbI$_3$ including both the tetragonal structure (160 – 330 K) and the pseudocubic structure (>330 K) using the first-principles-based atomistic simulations. Our simulation not only confirmed the very low thermal conductivity of the tetragonal MAPbI$_3$ at room temperature but also predicted a higher thermal conductivity in the pseudocubic phase. Thermal conductivity of MAPbI$_3$ shows a sudden jump from 0.57 W/m·K to 1.80 W/m·K from the tetragonal structure to the cubic structure at 330 K. The lower phonon group velocities and the stronger anharmonicty are found to be responsible for the lower thermal conductivity of the tetragonal MAPbI$_3$ than its pseudocubic counterpart.

**Acknowledgements**: This work was supported by the NSF (Grant No. 1512776). This work utilized the Janus supercomputer, which is supported by the National Science Foundation (Grant No. 0821794) and the University of Colorado at Boulder.

Table 1. Relaxed lattice structure obtained from DFT and MD simulations, compared with X-ray diffraction (XRD) results[15].

| Phases | $a$ (Å) | $b$ (Å) | $c$ (Å) | Cutoff (Ry) | k-mesh |
|---|---|---|---|---|---|
| Tetragonal | | | | | |
| DFT | 8.9405 | 8.7948 | 12.3937 | 60 | 6×6×4 |
| MD | 8.7790 | 8.7790 | 12.8816 | | |
| XRD | 8.8492 | 8.8492 | 12.6422 | | |
| Pseudocubic | | | | | |
| DFT | 6.2822 | 6.2298 | 6.3702 | 60 | 8×8×8 |
| MD | 6.3176 | 6.3066 | 6.3754 | | |
| XRD | 6.3115 | 6.3115 | 6.3161 | | |

Table 2. Potential field of the tetragonal and pseudocubic phases of MAPbI$_3$

| Interaction | Functional form | Parameters Tetragonal | Parameters Pseudocubic |
|---|---|---|---|
| **Pairwise:** | | | |
| Pb – I1 | Morse: $E = D[1 - e^{-\alpha(r-r_0)}]^2$ Units: $D$ in eV, $\alpha$ in Å$^{-1}$, $r$ and $r_0$ in Å | $D = 1.2253, \alpha = 0.7374, r_0 = 3.2233$ | $D = 1.2253, \alpha = 0.7374, r_0 = 3.171$ |
| Pb – I2 | | $D = 1.2253, \alpha = 0.7374, r_0 = 3.2260$ | $D = 1.2253, \alpha = 0.7374, r_0 = 3.171$ |
| Pb – I3 | | $D = 1.2253, \alpha = 0.7374, r_0 = 3.2233$ | $D = 1.2253, \alpha = 0.7374, r_0 = 3.171$ |
| C – N | | $D = 2.9721, \alpha = 2.0595, r_0 = 1.4990$ | |
| C – H1 | | $D = 3.6885, \alpha = 2.0055, r_0 = 1.0676$ | |
| C – H2 | | $D = 3.6851, \alpha = 2.1470, r_0 = 1.0119$ | |
| Pb – Pb | Buckingham [b]: $E = A\exp(-r/\rho) - \dfrac{C}{r^6}$, $r < 8$ Å Units: $A$ in eV, $\rho$ and $r$ in Å, $C$ in eV·Å$^6$ | $A = 3025475.81, \rho = 0.131258, C = 0$ | |
| I – I [a] | | $A = 980.1393, \rho = 0.282217, C = 29.96883$ | |
| Pb – C | | $A = 1405686.81, \rho = 0.150947, C = 0$ | |
| Pb – N | | $A = 1405686.81, \rho = 0.150947, C = 0$ | |
| I – C | | $A = 4860.90, \rho = 0.342426, C = 0$ | |
| I – N | | $A = 4860.90, \rho = 0.342426, C = 0$ | |
| C – C | Lennard – Jones: $E = 4\epsilon\left[\left(\dfrac{\sigma}{r}\right)^{12} - \left(\dfrac{\sigma}{r}\right)^6\right]$, $r < 8$ Å Units: $\epsilon$ in eV, $\sigma$ and $r$ in Å | $\epsilon = 0.004704, \sigma = 6.4386$ | $\epsilon = 0.004704, \sigma = 6.523$ |
| C – N | | $\epsilon = 0.005862, \sigma = 3.7538$ | $\epsilon = 0.005862, \sigma = 4.494$ |
| N – N | | $\epsilon = 0.007310, \sigma = 6.4386$ | $\epsilon = 0.007310, \sigma = 6.523$ |
| **Angular:** | | | |
| I1–Pb–I1 | Cosine – squared: $E = K(\cos\theta - \cos\theta_0)^2$ Units: $K$ in eV, $\theta$ and $\theta_0$ in degrees | $K = 0.8683, \theta_0 = 180$ | $K = 0.8683, \theta_0 = 170.2$ |
| I2–Pb–I2 | | $K = 0.8683, \theta_0 = 180$ | $K = 0.8683, \theta_0 = 171.8$ |
| I3–Pb–I3 | | $K = 0.8683, \theta_0 = 180$ | $K = 0.8683, \theta_0 = 164.6$ |
| I1–Pb–I2 | | $K = 0.8683, \theta_0 = 90$ | $K = 0.8683, \theta_0 = 90$ |
| I1–Pb–I3 | | $K = 0.8683, \theta_0 = 90$ | $K = 0.8683, \theta_0 = 90$ |
| I2–Pb–I3 | | $K = 0.8683, \theta_0 = 90$ | $K = 0.8683, \theta_0 = 90$ |
| Pb–I1–Pb | | $K = 0.8683, \theta_0 = 180$ | $K = 0.8683, \theta_0 = 179.7$ |
| Pb–I2–Pb | | $K = 0.8683, \theta_0 = 163.7$ | $K = 0.8683, \theta_0 = 171.8$ |
| Pb–I3–Pb | | $K = 0.8683, \theta_0 = 163.7$ | $K = 0.8683, \theta_0 = 164.7$ |
| N–C–H1 | | $K = 1.5282, \theta_0 = 109.2$ | |
| H1–C–H1 | | $K = 1.5872, \theta_0 = 110.3$ | |
| C–N–H2 | | $K = 1.5426, \theta_0 = 112.3$ | |
| H2–N–H2 | | $K = 1.5491, \theta_0 = 106.4$ | |
| **Coulombic [c]:** | | $\epsilon_r = 5.672$.[d] | |
| | | Bader charges [e]: | |
| Pb | $E = \dfrac{Cq_iq_j}{\epsilon_r r}, r < 8$ Å $C$ is a unit-conversion constant. $q_i$ and $q_j$ are multiples of the elementary charge. | 0.9155 | 0.9150 |
| I1 | | -0.5492 | -0.5731 |
| I2 | | -0.5538 | -0.5664 |
| I3 | | -0.5678 | -0.5553 |
| C | | 0.5203 | 0.4661 |
| N | | -1.4813 | -1.4713 |
| H1 | | 0.0699 | 0.0920 |
| H2 | | 0.5022 | 0.5030 |

a. An element symbol without numbers denotes any atom type of this element. For example, I-I means the interaction between any two Iodine atoms.

b.  The parameters of the Buckingham functions are optimized based on ref. [37]
c.  Coulombic interactions are not switched off between any pairs with interatomic distance smaller 8 Å.
d.  The dielectric constant is obtained by minimizing the force deviation function $W$ in equation (1). The value is found close to the high-frequency optical dielectric constant (5.6 – 6.5) [38, 39] of MAPbI$_3$.
e.  Bader charges are fixed during the minimization of the force deviation function $W$.

Table 3. Acoustic velocities calculated from the phonon dispersion.

| Phases | Direction | LA (m/s) | TA1 (m/s) | TA2 (m/s) |
|---|---|---|---|---|
| Tetragonal | | | | |
| | [001] | 1963 | 927 | 1199 |
| | [100] | 2239 | 1236 | 1236 |
| Pseudocubic | | | | |
| | [100] | 2824 | 1314 | 1314 |

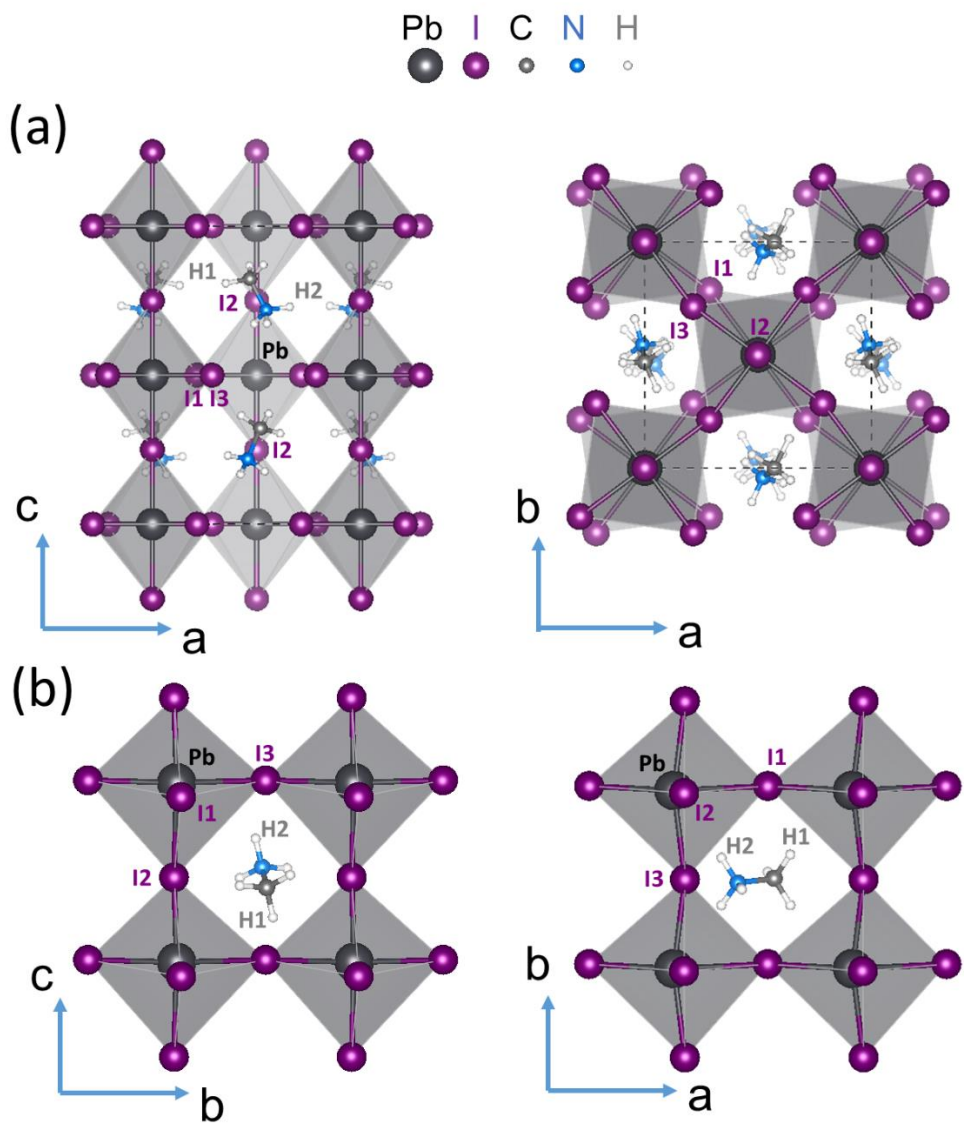

Figure 1. The crystal structure of (a) the tetragonal and (b) the pseudocubic phases of MAPbI$_3$ projected in (010) and (001) plane from left to right correspondingly. The a and b axes of the both phases are defined along the two directions with smaller lattice constants, and the c axis is defined along the direction with the largest lattice constant. The legend on the top indicates different elements depicted in the figure. A more detailed atomic types used in the potential field are also indicated with element symbols followed by numbers. H1 and H2 are hydrogen atoms connected to the C atom and N atom, respectively.

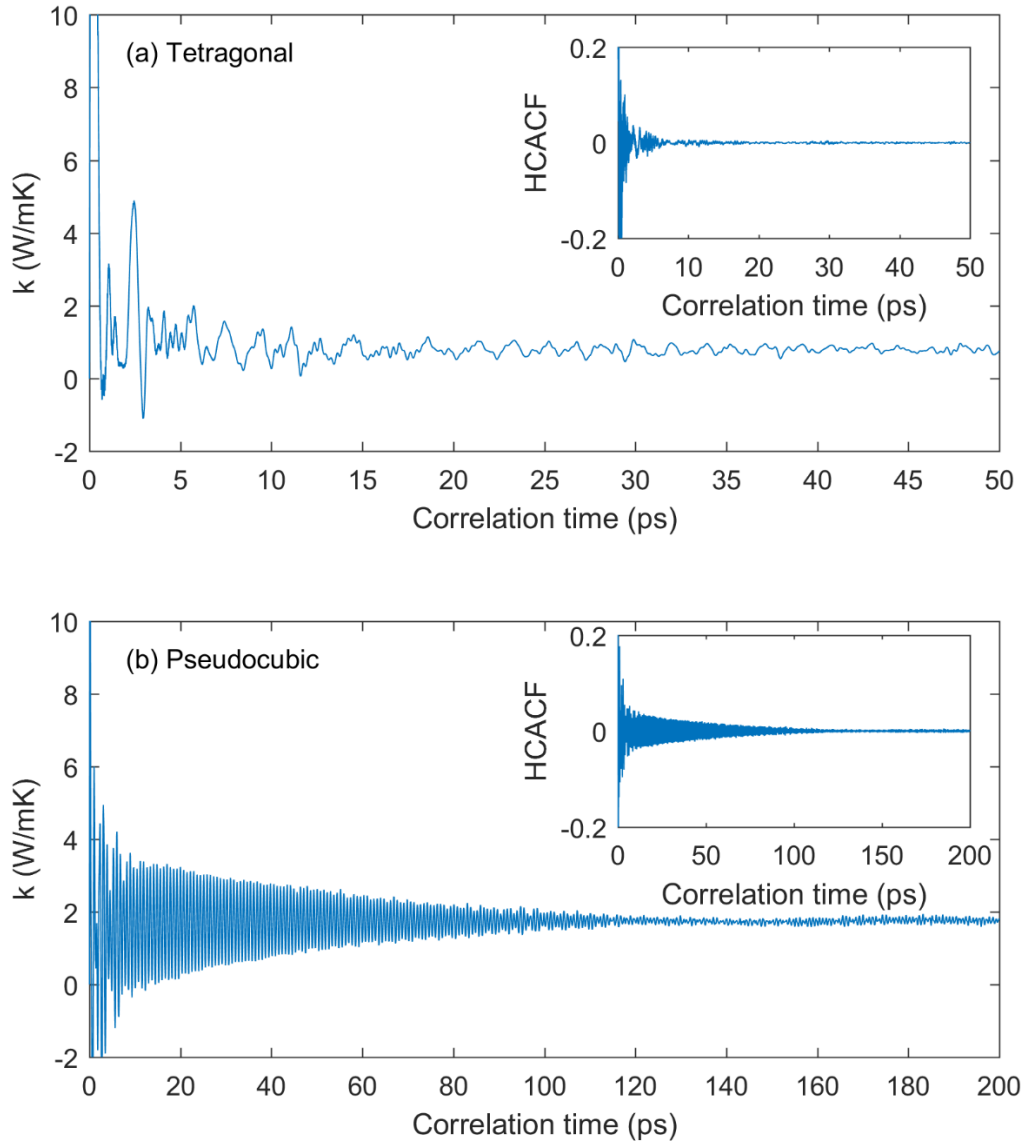

Figure 2. The thermal conductivity of (a) the tetragonal and (b) the pseudocubic MAPbI$_3$ obtained by Green-Kubo relation plotted against correlation time. The insets show the decay of normalized heat current autocorrelation function (HCACF).

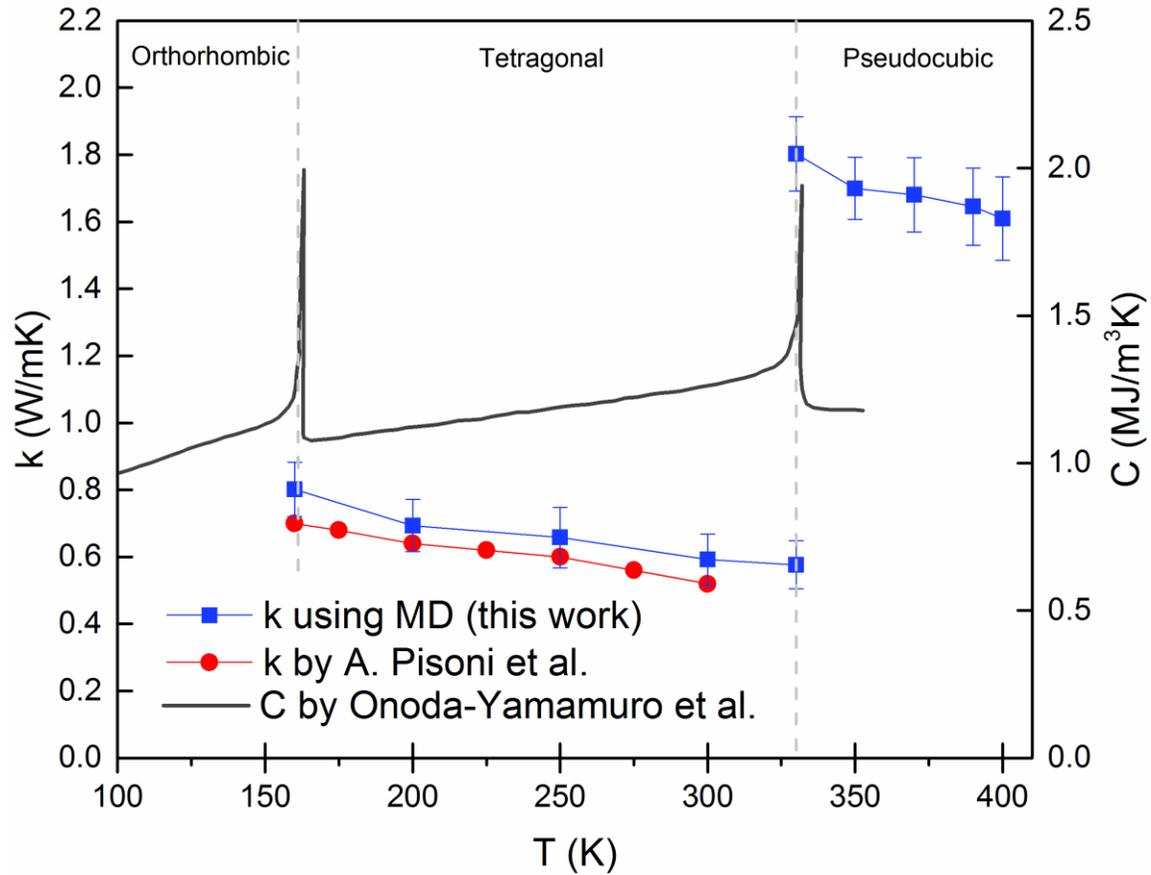

Figure 3. Temperature-dependent thermal conductivity of MAPbI$_3$ from 160 K to 400 K, compared with the experimental results by A. Pisoni et al..[14] The volumetric heat capacity based on the measurement by Onoda-Yamamuro et al.[31] is also plotted as the right Y axis, whose peaks indicate phase change.

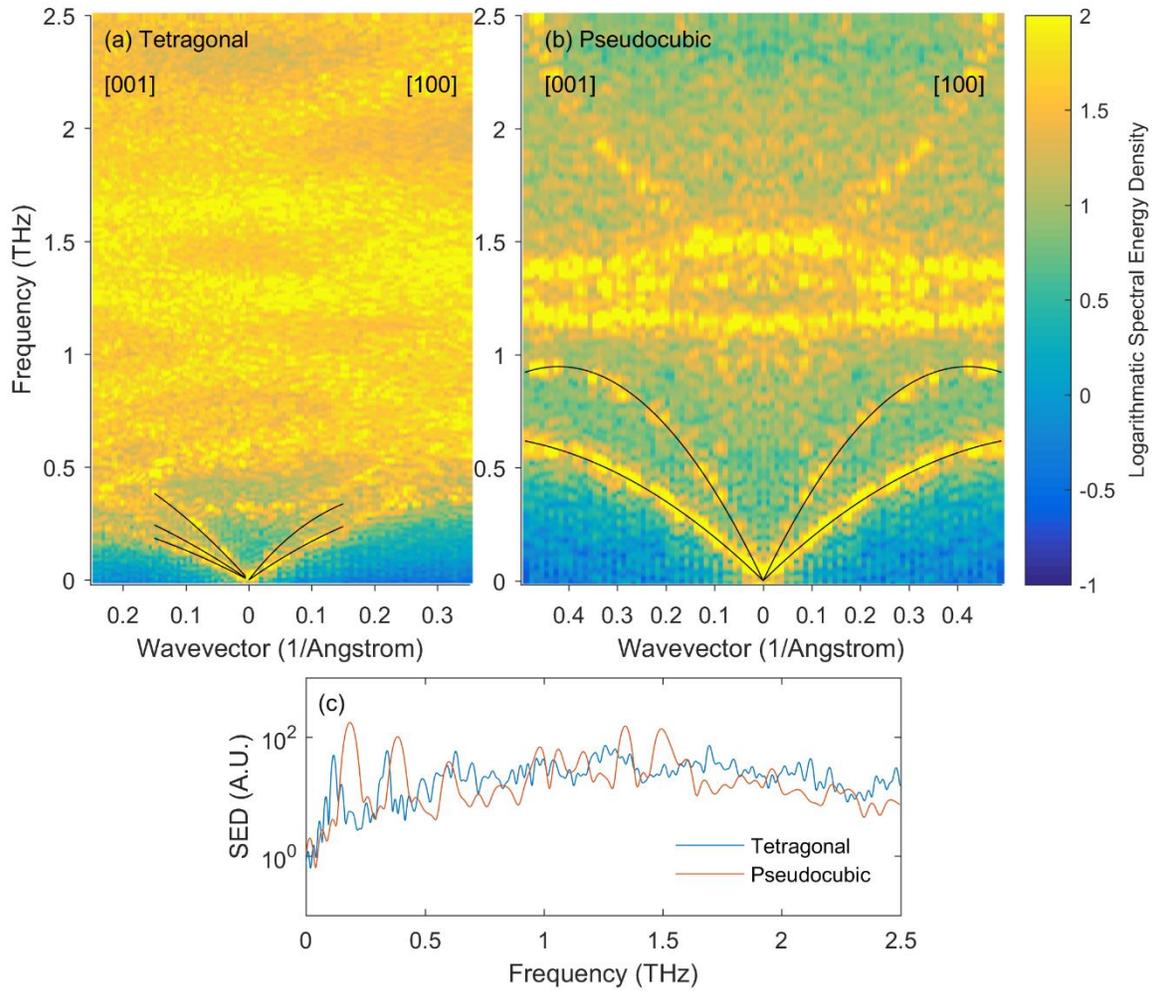

Figure 4. Phonon dispersion of the (a) tetragonal and (b) pseudocubic MAPbI$_3$ by the SED method. (c) The SED peaks of both phases at wavevector (0.1,0,0).

# Supplementary Information

Lattice Thermal Conductivity of Organic-Inorganic Hybrid Perovskite $CH_3NH_3PbI_3$

Xin Qian, Xiaokun Gu and Ronggui Yang[*]

Department of Mechanical Engineering & Materials Science and Engineering Program,

University of Colorado, Boulder, Colorado 80309-0427, USA

[*]E-mail: Ronggui.Yang@Colorado.edu

Table of Contents

# Supplementary Information

**S1. Convergence test of k-mesh density and energy cutoff in DFT calculations**

We have performed a series of DFT calculations of ground-state energy by increasing the k-mesh density and energy cutoff, as shown in Figure S1. For the tetragonal phase, convergence of ground-state energy is achieved when the k-mesh is denser than 6×6×4. For the pseudocubic phase, convergence is achieved when the k-mesh is denser than 8×8×8. Figure S2 shows convergence of ground-state energy in respect energy cutoff, with a 6×6×4 k-mesh density for the tetragonal phase and 8×8×8 k-mesh density for the pseudocubic phase. For both phases, ground-state energy is converged when using a cutoff larger than 60 Ry. In this paper, we have thus used 60 Ry as the energy cutoff to present the results.

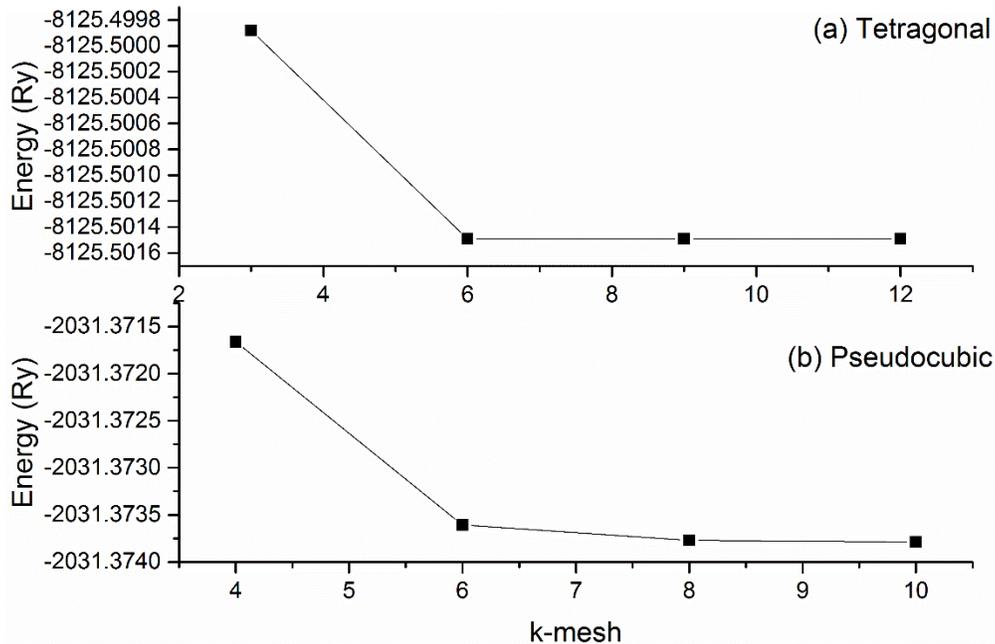

Figure S1. Convergence of ground-state energy with k-mesh density in the a-axis of the crystal. In tetragonal phase, the number ratio of k-points in three directions is kept at 3:3:2. In the pseudocubic phase, the number of k-points is equal in each direction.

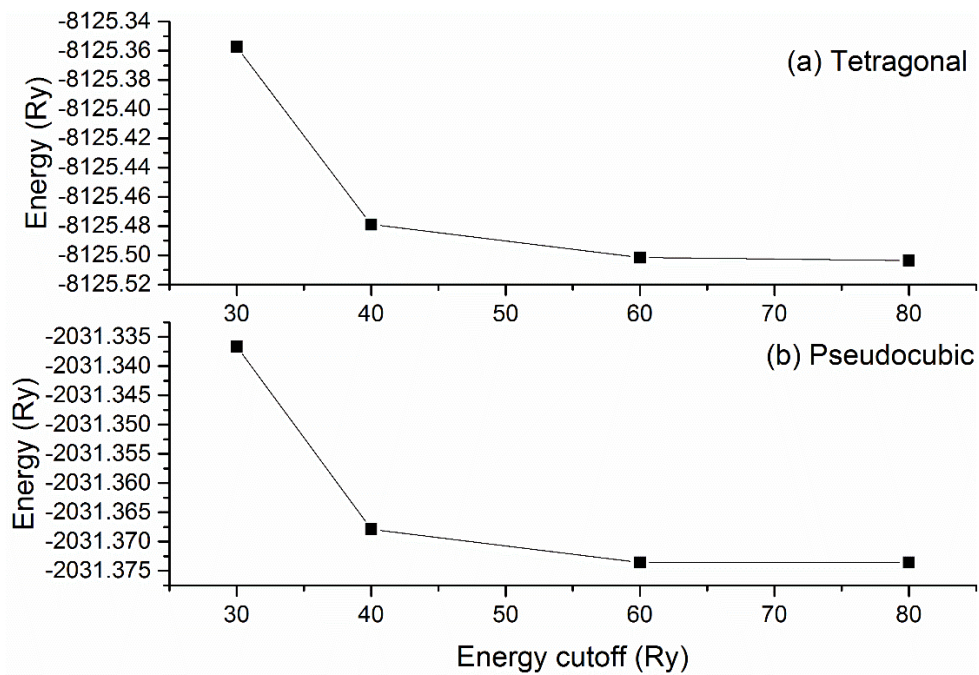

Figure S2. Convergence of ground-state energy with energy cutoff.

**S2. Effect of simulation domain size on thermal conductivity calculations**

We studied the dependence of calculated thermal conductivity of MAPbI$_3$ on the size of simulation domain, as shown in Figure 3.[1] For the tetragonal phase, the thermal conductivities based on different simulation domain sizes are calculated at 160 K. Five runs with different initial atomic velocities are performed independently, and the final result is averaged. Convergence is observed when the simulation cell is larger than 5×5×5 unit cells (Figure S3a). Since the phonon mean free paths are expected to be smaller at higher temperatures, a simulation cell containing 6×6×6 unit cells should be large enough at temperature higher than 160 K for the tetragonal phase. Similarly, we studied the dependence of calculated thermal conductivity of the pseudocibic MAPbI$_3$ on the size of simulation domain at 330 K. Figure S3b shows that thermal conductivity is converged when using a 6×6×6 simulation cell size. We have thus chosen a simulation cell size of 6×6×6 unit cells for both the tetragonal and pseudocubic phase to obtain and present the data in the letter.

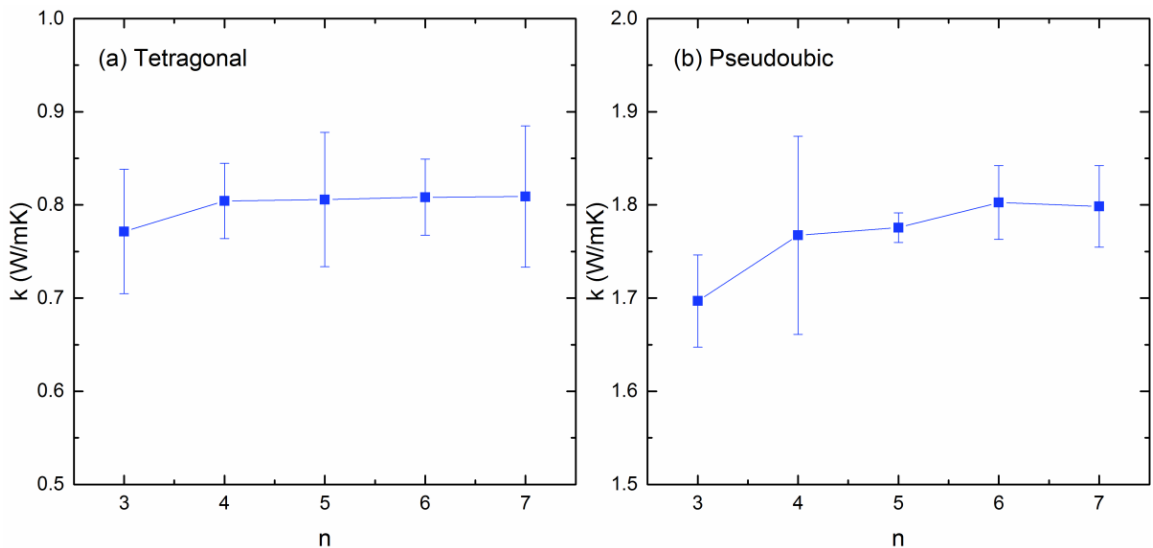

Figure S3. Size effect of thermal conductivity in MAPbI$_3$ for the tetragonal (160 K) and the pseudocubic phase (330 K). The simulation cell contains n×n×n unit cells.

**S3. Accuracy of the potential field**

We have calculated the phonon density of states (DOS) MAPbI$_3$ using MD simulation as shown in Figure S4, compared with the DFT calculation by Brivio et al.[2] Qualitatively, our potential field has reproduced similar vibrational properties with four energetic regions of phonons: (i) a low frequency band from 0–7 THz mainly comes from the vibration of PbI$_6$ octahedrons (ii) an isolated peak near 9 THz corresponds to the twist of MA cation around C-N bond (iii) a mid-frequency band from 20–50 THz, and (iv) a high-frequency band from above 80 THz come from the vibration of the organic cations. Generally, the DOS peaks by MD calculation is broadened compared to the DFT calculation because the MD calculation is intrinsically at an elevated temperature at 300 K for the tetragonal phase and 350 K for the pseudocubic phase. The peaks in the mid-frequency

# Supplementary Information

band by our MD calculation are closer to each other compared to DFT calculation, but the frequency difference is still within 20%. Further improving the matching of the phonon spectrum is very challenging due to the complexity of the crystal structure.

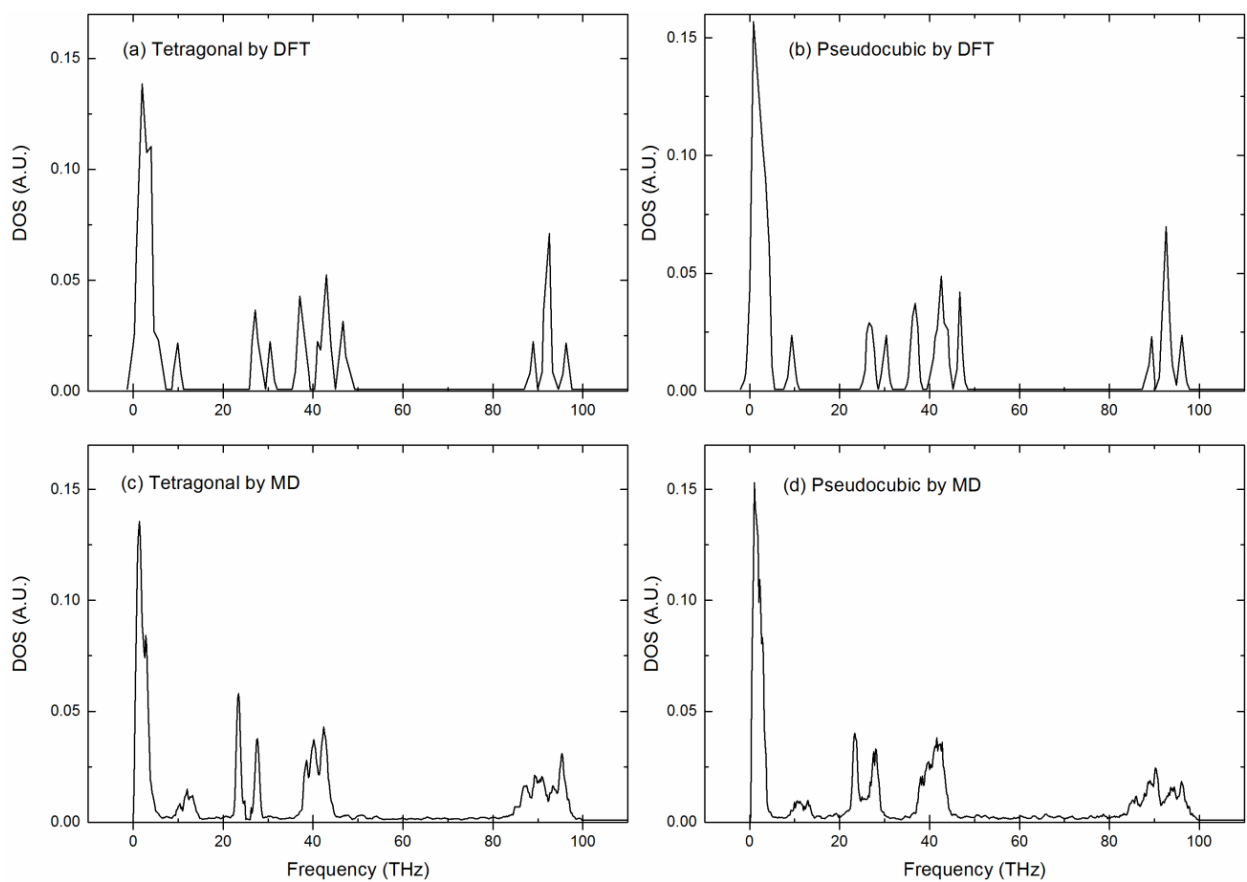

Figure S4. Phonon density of states (DOS) of MAPbI$_3$ in the tetragonal phase and pseudocubic phase, compared with DFT calculations in Ref [2].

**References**

1. X. Li, K. Maute, M. L. Dunn and R. Yang, Phys. Rev. B **81** (24) (2010).
2. F. Brivio, J. M. Frost, J. M. Skelton, A. J. Jackson, O. J. Weber, M. T. Weller, A. R. Goñi, A. M. A. Leguy, P. R. F. Barnes and A. Walsh, Phys. Rev. B **92** (14) (2015).